\documentclass{article}
\pdfoutput=1
\usepackage{jcappub}

\usepackage{bm}

\newcommand{\be}{\begin{equation}}
\newcommand{\ee}{\end{equation}}
\newcommand{\ba}{\begin{eqnarray}}
\newcommand{\ea}{\end{eqnarray}}

\title{Testing homogeneity with the fossil record of galaxies}

\author[a]{Alan F. Heavens,}
\author[b]{Raul Jimenez}
\author[c,d]{and Roy Maartens}

\affiliation[a]{SUPA, Institute for Astronomy, University of Edinburgh, Blackford Hill, Edinburgh, EH9 3HJ, UK}
\affiliation[b]{ICREA \& ICC, University of Barcelona (IEEC-UB), Barcelona 08024, Spain}
\affiliation[c]{Department of Physics, University of Western Cape, Cape Town 7535, South Africa}
\affiliation[d]{Institute of Cosmology \& Gravitation, University of Portsmouth, Portsmouth~PO1~3FX, UK}

\emailAdd{afh@roe.ac.uk}
\emailAdd{raul.jimenez@icc.ub.edu}
\emailAdd{roy.maartens@port.ac.uk}

\abstract{
The standard Friedmann model of cosmology is based on the Copernican Principle, i.e. the assumption of a homogeneous background on which structure forms via perturbations. Homogeneity underpins both general relativistic and modified gravity models and is central to the way in which we interpret observations of the CMB and the galaxy distribution. It is therefore important to probe homogeneity via observations. We describe a test based on the fossil record of distant galaxies: if we can reconstruct key intrinsic properties of galaxies as functions of proper time along their worldlines, we can compare such properties at the same proper time for our galaxy and others.  We achieve this by computing the lookback time using radial Baryon Acoustic Oscillations,  and the time along galaxy world line using stellar physics, allowing us to probe homogeneity, in principle anywhere inside the past light cone.  Agreement in the results would be an important consistency test -- although it would not in itself prove homogeneity. Any significant deviation in the results however would signal a breakdown of homogeneity.
}


\begin{document}

\maketitle

%%%%%%%%%%%%%%%%%%%%%%%%%%%%%%%%%%%%%%%%%%%%%%%%%%%%%%%%%%%%%%%%%%%%%%%
%%%%%%%%%%%%%%%%%%%%%%%%%%%%%%%%%%%%%%%%%%%%%%%%%%%%%%%%%%%%%%%%%%%%%%%
\section{Introduction}

The standard model of the Universe is homogeneous, with structure formation described via perturbations on a fixed background. Within general relativity (GR), it then follows that the acceleration of the Universe is driven by dark energy. The homogeneous standard model is a simple, predictive model that successfully accommodates all observations up to now. However, we should probe the foundations of this model as far as possible. This is further motivated by the fact that there is still no satisfactory description of the dark energy that is central to the model. We can probe the assumption that GR holds on cosmological scales, by investigating modified gravity theories and by devising consistency tests of GR. This probe is only effective if we assume homogeneity. Alternatively, we can probe the assumption of homogeneity itself \cite{Maa11,Ellis:2011hk,ClaMaa10}.

Homogeneity is {\em not} established by observations of the CMB and the galaxy distribution -- {\em we cannot directly observe homogeneity,} since we observe down the past lightcone, recording properties on 2-spheres of constant redshift and not on spatial surfaces that intersect that lightcone. What these observations can directly probe is isotropy. In order to link isotropy to homogeneity, we have to assume the Copernican Principle, i.e. that we are not at a special position in the Universe. The Copernican Principle is not observationally based; it is an expression of the intrinsic limitation of observations from one spacetime location.

We cannot observationally prove homogeneity, but we can develop consistency tests that would uncover deviations from homogeneity if these deviations existed. In other words, we can test for consistency of the Copernican Principle and for violations of the Copernican Principle.  If we find no violation, then the indirect evidence for homogeneity is strengthened. However,  any single significant violation of the Copernican Principle would disprove homogeneity.

Here we propose a new test of homogeneity, based on intrinsic properties of galaxy evolution. Homogeneity means that fundamental properties, such as star formation histories, should be uniform -- this is the `fossil record' implicitly carried by galaxies via their evolution. By observing distant galaxies, and then tracing our own galactic history back to the proper time of emission, we can check for consistency between properties of our galaxy and distant galaxies.

This is closely related to the `cosmic chronometer' project \cite{Stern:2009ep,Crawford:2010rg,Moresco:2010wh}, which uses galaxy spectral properties to find $H(z)$ in a Friedmann model 
%*
(see also \cite{Carson:2010sq}). Here we do not assume Friedmann geometry but rather test for it. It is also related to earlier work which attempted to prove homogeneity from uniformity of the thermal history of galaxies \cite{bonell86}. However it turns out that there are inhomogeneous models with uniform thermal histories \cite{bonell86}, and the proposal fails as a proof of homogeneity. It does survive however as a probe of homogeneity -- i.e., violations of uniformity would indicate a breakdown of the Copernican Principle and thus of homogeneity. The question is how to apply the idea in practice.

Before describing our proposal, we briefly review other tests of the Copernican Principle and homogeneity.

%\section{TESTING THE CP AND HOMOGENEITY}\label{cpsec}

%Various tests of homogeneity have been proposed.

\begin{itemize}
\item

The geometric null test of
\cite{CBL} is based on the tight relation between curvature and expansion history in a Friedmann spacetime. This relation is independent of dark energy and other contributions to the energy-momentum tensor (and also independent of the field equations). The quantity
\begin{eqnarray}
{\cal K}(z) &:=& H_0^2(1+z)^4 + H^2(z)\left[(1+z)^2\left\{D_L(z)D_L''(z)- D_L^{\prime 2}(z)\right\}+D_L^2(z) \right]
\nonumber\\ &&~
+ (1+z)H(z) H'(z)D_L(z)\left[(1+z) D_L'(z)-D_L(z) \right]  \label{cz test}
\end{eqnarray}
is identically zero at all redshifts in a Friedmann spacetime with arbitrary content and curvature.
It follows that
 \be
{\cal K}(z) ~~\mbox{significantly different from 0} ~~\Rightarrow %*~~ \mbox{violation of CP and homogeneity.}
~~ \mbox{violation of homogeneity.}
 \ee
(How to implement the test is discussed in \cite{Shafieloo:2009hi}.)

\item
The baryon acoustic oscillation (BAO) feature in the galaxy distribution provides a further geometric test of homogeneity %*\cite{CBL,Maa11}. 
\cite{CBL,Clarkson:2010ej,Maa11}.
Future large-volume surveys will allow the detection of the BAO scale in both radial and transverse directions.
These scales are equal in a Friedmann background, so that
 \be
{\mbox{radial BAO} \over \mbox{transverse BAO}}-1  ~~\mbox{significantly different from 0} ~~\Rightarrow 
%*~~ \mbox{violation of CP and homogeneity.}
~~ \mbox{violation of homogeneity.}
 \ee

\item

Galaxy clusters with their hot ionized intra-cluster gas, act via scattering of CMB photons like giant mirrors that carry information about the last scattering surface seen by the cluster -- and therefore serve as indirect probes inside our past lightcone.
The Sunyaev-Zeldovich (SZ) effect on the CMB temperature anisotropies probes the monopole (thermal SZ) and dipole (kinetic SZ) seen by the cluster in a perturbed Friedmann model. Thus \cite{goodman,Maa11}
 \be
\mbox{Non-perturbative thermal or kinetic SZ temperature effect}  ~~\Rightarrow
%*~~ \mbox{violation of CP and homogeneity.}
~~ \mbox{violation of homogeneity.}
 \ee
(In related work, the SZ effect has been used to place constraints on inhomogeneous void models %*\cite{CS,Zhang:2010fa}.)
\cite{CS,Zhang:2010fa,Moss:2011ze}.)

The SZ effect on CMB {\em polarization} provides a new and different probe of the monopole and dipole, and in addition probes the quadrupole and octupole seen by the cluster, in a perturbed Friedmann model. Thus \cite{Maa11}:
 \be
\mbox{Non-perturbative SZ polarization effects}  ~~\Rightarrow 
%*~~ \mbox{violation of CP and homogeneity.}
~~ \mbox{violation of homogeneity.}
 \ee

\end{itemize}
Other tests that may become possible with future observations include: a geometric null test based on the time drift of the cosmological redshift \cite{UCE}; massive neutrinos propagate inside our past lightcone and provide in principle a probe of deviations from homogeneity \cite{Jia:2008ti}, if the cosmic neutrino background can be observed.

We now turn to our proposed probe of the Copernican Principle, based on the fossil record carried by galaxies, which is also an indirect probe of conditions inside our past lightcone. First we must deal with the problem of lookback time.

\section{Lookback time in a general universe}

\begin{center}
\begin{figure}
\hspace*{4cm}
\includegraphics[width=.5\columnwidth]{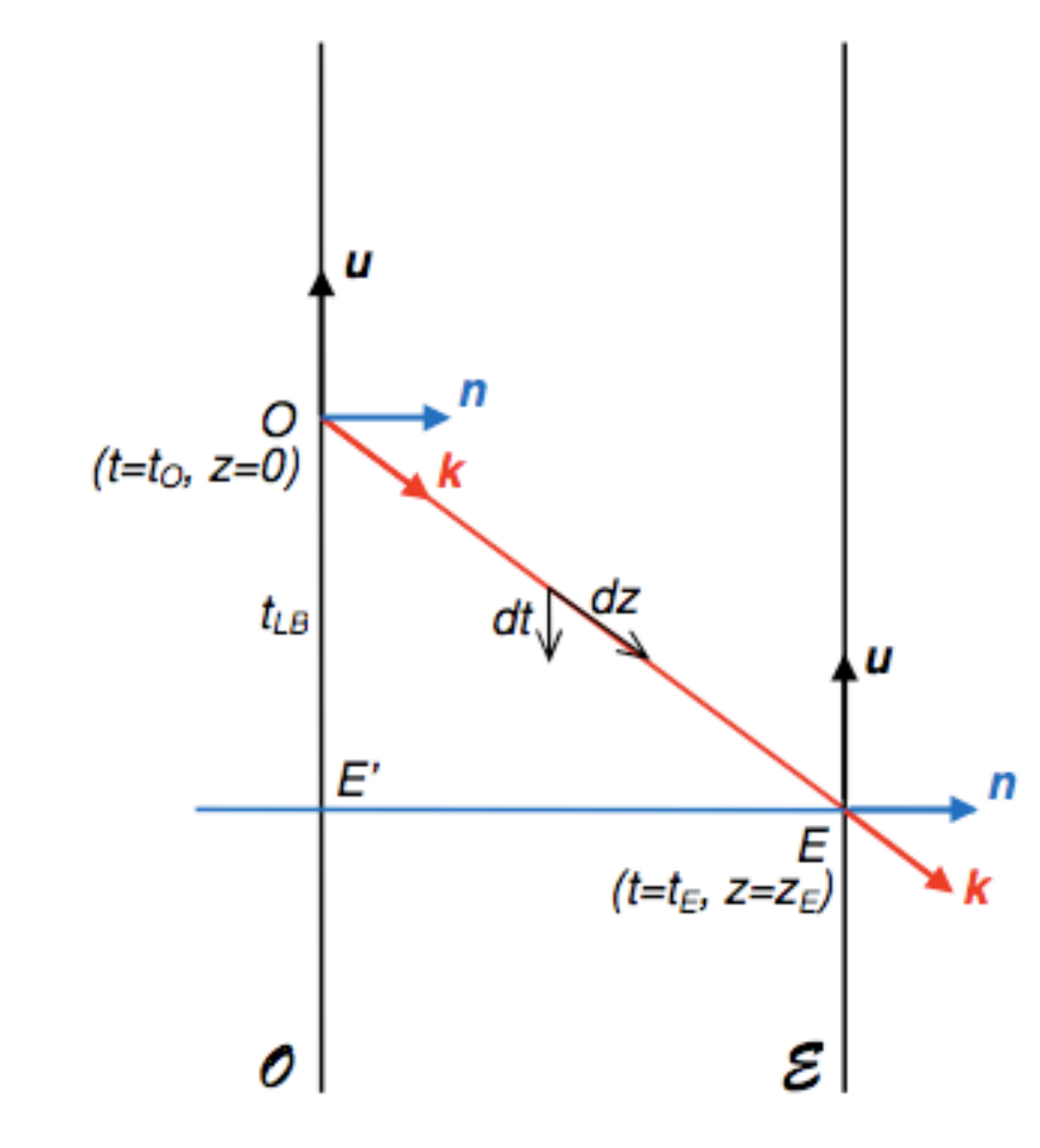}
\caption{Lookback time in a general spacetime.}\label{fig-lb}
\end{figure}
\end{center}

A distant galaxy, with worldline ${\cal E}$, emits photons at event $E$ that we observe with redshift $z_E$ at event $O$ on our galaxy worldline ${\cal O}$ (see Fig. \ref{fig-lb}).
In order to compare the intrinsic properties of ${\cal E}$ and ${\cal O}$ at the same proper time, we need to compute the lookback time $t_{LB}=t_O-t_E$, where
$t$ denotes proper time along galaxy worldlines. This is straightforward in a Friedmann model -- but we cannot assume the geometry of spacetime if our aim is to test for homogeneity. So we need to compute the lookback time in a covariant way, valid in a general spacetime.

The galaxy 4-velocity field is $u^\mu=dx^\mu/dt$. The past-pointing photon 4-momentum is $k^\mu=dx^\mu/dv$, where $v$ is the null affine parameter with $v=0$ at $O$. Then
 \be \label{zkdef}
1+z=u_\mu k^\mu, ~~ k^\mu=(1+z)(-u^\mu+n^\mu), ~~ u_\mu n^\mu=0,~ n_\mu n^\mu=1,
 \ee
where $n^\mu$ is a unit vector along the line of sight. For observers comoving with the matter, an increment $dv$ in null affine parameter corresponds to a time increment $dt$, where
 \be \label{tvz}
dt=-u_\mu k^\mu dv=-(1+z)dv.
 \ee

We need to relate $v$ to $z$: by (\ref{zkdef}),
 \be \label{dzdvgen}
{dz \over dv}= k^\nu \nabla_\nu (u_\mu k^\mu)= k^\mu k^\nu \nabla_\mu u_\nu ,
 \ee
where the last equality follows since $k^\mu$ is geodesic. The covariant derivative is split as
 \be
\nabla_\mu u_\nu={1\over 3}\Theta h_{\mu\nu}+ \sigma_{\mu\nu}+\omega_{\mu\nu}-u_\mu \dot{u}_\nu,
 \ee
where $h_{\mu\nu}=g_{\mu\nu}+u_\mu u_\nu$ projects into the galaxy instantaneous rest space, $\Theta$ is the volume expansion rate ($\Theta=3H$ in a Friedmann model), $\sigma_{\mu\nu}$ is the shear, $\omega_{\mu\nu}$ is the vorticity and $\dot{u}_\mu$ is the acceleration. Since matter is pressure-free, $\dot{u}_\mu=0$. Putting everything together, we get
 \be \label{dzdv}
{dz \over dv}=(1+z)^2 \Big[{1\over 3}\Theta
%-\dot{u}_\mu n^\mu
+ \sigma_{\mu\nu} n^\mu n^\nu \Big].
 \ee

Now we integrate along the lightray from $O$ to $E$, using (\ref{tvz}) and (\ref{dzdv}):
 \ba \label{lbt}
t_O-t_E= \int_0^{z_E} {dz \over (1+z) \Big[ \Theta(z)/3 + \sigma_{\mu\nu}(z) n^\mu n^\nu \Big]}.
 \ea
This will give us the lookback time -- {\em provided that we can uniquely relate the time intervals along galaxy worldlines that cross the light ray to a time interval along our worldline} ${\cal O}$. In order to do this, we need the existence of spatial 3-surfaces that are everywhere orthogonal to $u^\mu$; these will then be surfaces of constant proper time. The necessary and sufficient condition for these surfaces to exists is an irrotational flow:
 \be
\omega_{\mu\nu}=0.
 \ee
Then we can uniquely identify the event $E'$ where the constant proper time surface $t=t_E$ through $E$ intersects ${\cal O}$. (For rotating matter, it is not clear whether we can consistently define a lookback time.)

\section{Probing homogeneity with the fossil record}

In order to compare the two galaxies at the same proper time, we need a method to evaluate the lookback time (\ref{lbt}) from some observable. The observed expansion rate along the line of sight is \cite{ClaMaa10}
 \be
H_{\rm obs, los}(z)=n^\mu n^\nu\nabla_\mu u_\nu  = {1\over 3}\Theta(z) + \sigma_{\mu\nu}(z) n^\mu n^\nu.
 \ee
This leads to a covariant expression for the radial part of the Alcock-Paczynski formula, valid in any spacetime; in particular, we get a covariant expression for the comoving radial BAO scale, set by the maximum comoving distance a sound wave travels, $r_{\rm bao,los}$, without assuming Friedmann geometry \cite{Maa11}:
 \be \label{rbao}
r_{\rm bao, los} ={ \delta z(z) \over H_{\rm obs, los}(z)}={ \delta z(z) \over \Theta(z)/3 + \sigma_{\mu\nu}(z) n^\mu n^\nu}.
 \ee
In principle, $ H_{\rm obs, los}$ can be found from observations of the extent of the baryon ruler in redshift, $\delta z(z)$, for a range of redshifts along the line of sight to the emitting galaxy. Then the lookback time is simply
 \be\label{lbobs}
t_O-t_E= \int_0^{z_E} {dz \over (1+z)H_{\rm obs, los}(z)}.
 \ee

We can estimate how many independent patches there are in the sky where the BAO feature can be measured, using the Friedmann formalism in \cite{seo}.  Note that radial BAO observations give us $H_{\rm obs,los}$ times the $r_{\rm bao,los}$; we assume the latter is computed to sufficient accuracy ($\sim 1\%$ in the standard model) that the main error source is from $H_{\rm obs, los}$. The angular size of a box with redshift depth of $\delta z$ is $\delta\theta=(1+z)D_A/ r_{\rm bao, trans}$, where for this estimate we assume $ r_{\rm bao, los}= r_{\rm bao, trans}$. Then taking redshift bins of, for example,  width $0.2$, a
10 (20)\% measurement of  $H_{\rm obs, los}(z)$ requires boxes with size $\sim 650 (450)\mbox{Mpc}h^{-1}$. We ignore complications from bias and assume that the survey will be large enough such that shot noise is subdominant. This estimate is supported by looking directly into a suite of dark matter simulations and determining the BAO feature \cite{wagner}. Therefore, there are $\sim 50$ independent patches in the sky where the homogeneity test can be performed.

For the galaxy fossil record, we assume that from the integrated spectrum of a galaxy ${\cal E}$ observed at redshift $z_E$, we can obtain its specific star formation rate $S$
at the event $E$ of emission, corresponding to ${\cal E}$'s proper time $t_E$ (see Fig. \ref{fig-lb}).
Then we use a model of star formation to compute $S$ for our galaxy in the past,  at the event $E'$, corresponding to proper time $t=t_O-t_E$, which we determine from the radial BAO data via (\ref{lbobs}).
If the universe is homogeneous, then these rates should agree within the errors -- and this should be true for galaxies observed in any direction and at any redshift where we are able to determine the lookback time. Conversely, any significant difference for even one comparison would indicate a breakdown of homogeneity:
 \be
S_{E'}(t_O-t_E)-S_E(t_E) ~~\mbox{significantly different from 0} ~~\Rightarrow 
%*~~ \mbox{violation of CP and homogeneity.}
~~ \mbox{violation of homogeneity.}
 \ee

A more ambitious approach is to obtain the lookback time also from stellar ages of the stellar populations of galaxies.
If the Universe is homogeneous, then in patches of the sky that are sufficiently large to measure global properties, $S$ should be be the same  {\em at the same time from the beginning of star formation in galaxies}.  We thus need both absolute ages and relative ones for the different time bins in which the galaxy lookback time has been split.

We obtain the above determination of the specific star formation rate and lookback time for two galaxies, our own ${\cal O}$ and ${\cal E}$. We also measure the redshift $z_E$. Let us consider the different possible outcomes. Suppose that galaxy ${\cal O}$ has an absolute age $T$. If we measure that galaxy ${\cal E}$ has an age $> T$ then we conclude that the universe is inhomogeneous, as it is otherwise impossible to explain how the distant galaxy with positive lookback time is older.

\begin{figure}
\begin{center}
\includegraphics[width=\columnwidth]{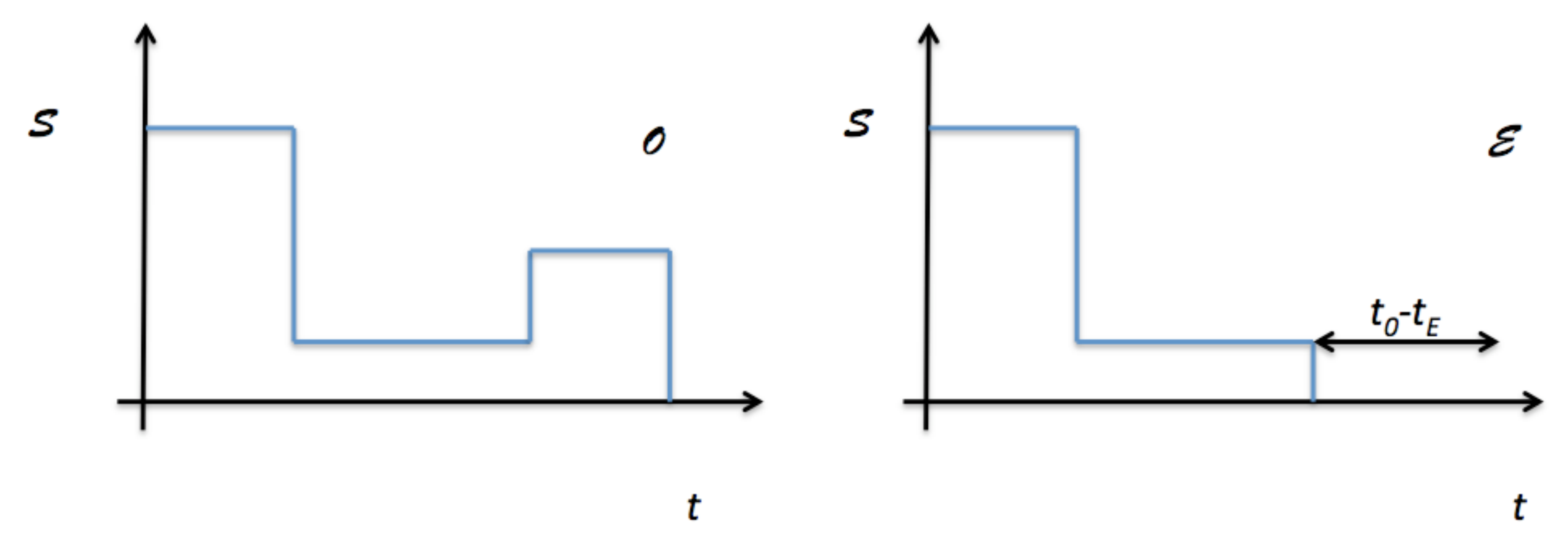}
\end{center}
\caption{Schematic of star formation histories for 2 galaxies showing how the time difference from the last period of activity can be used to probe homogeneity.}
\label{ssfr}
\end{figure}

The other outcome is that the absolute age of galaxy ${\cal E}$ is smaller than $T$. Then we need to look at the specific star formation rate per unit of mass for both galaxies. In a homogeneous Universe, these should be the same {\em at the same time from t=0}. Comparing the specific star formation rates, we look at the time difference from the last event when stars where formed at galaxy ${\cal E}$ and galaxy ${\cal O}$ (see Fig. \ref{ssfr}).
Because we know the redshift, we can now compare with the lookback time (\ref{lbobs}) determined via the BAO.

This method would not work if the specific star formation rate is an exact power law, but this is never the case \cite{Heavens04,Panter07}.
In order to perform this test in practice we can use the currently determined  MOPED/VESPA \cite{Heavens04,Tojeiro07} star formation histories. 
For an individual galaxy the time resolution is coarse (about 30\% of the Hubble time), but we can improve the {\em spatial} resolution of the test of homogeneity
by binning the spectra (cf. \cite{Tojeiro:2010up}), thus improving the {\em time} resolution of the recovered star formation histories.   
At some point the method is limited by the accuracy of current models, but 10\% resolution should be possible with current data and models, allowing a test of homogeneity on scales of a few hundred Mpc.  We are currently working on such a test (Hoyle et al., in preparation).

\section{Conclusions}\label{discon}

Although we cannot directly observe homogeneity, we can test the Copernican Principle at the foundation of homogeneity, using observations that carry information from inside our past lightcone. Previously this has been discussed via consistency relations between distances and expansion rates, using supernova and BAO data, and via Sunyaev-Zeldovich effects from galaxy clusters on the CMB temperature and polarization. Here we have proposed a new test of homogeneity based on the fossil record of distant galaxies. We make observations at constant time slices within our past light cone by computing the lookback time to a distant galaxy and adding it to the proper time along the galaxy past worldline.  The  lookback time is computed without assuming Friedmann geometry, using standard rulers (the radial BAO); the proper time on the galaxy worldline is effectively computed using the fossil record, dependent on the physics of stellar evolution but not on a Friedmann geometry.    

As remarked, this test has the advantage that it allows one to look {\em inside} our past light cone, in principle everywhere inside.  Naturally, we cannot make any global statements about homogeneity, as we are inevitably restricted to the region on or within our past light cone.  The method is interesting in the sense that once the lookback time has been computed, then the fossil record comparisons should be limited only by observational error and sample variance, and if fossil record observations are not consistent in different locations then homogeneity does not apply.  The converse is that any failure to find inhomogeneity strengthens the case for the Copernican Principle.

\[ \]{\bf Acknowledgments:}
We thank Chris Clarkson,  Catherine Cress, George Ellis, Licia Verde and Christian Wagner for discussions. RM is supported by a South African SKA Research Chair, by the UK Science \& Technology Facilities Council (grant no. ST/H002774/1) and by a NRF (South Africa)/ Royal Society (UK) exchange grant.

\end{document}